\documentclass[useAMS,usenatbib]{mn2e}
\usepackage{psfig}
\usepackage{url}
\usepackage[dvips]{graphicx}
\def\mnras{{MNRAS}}
\def\apj{{ApJ}}

\def\aap{{A\&A}}
\def\apjl{{ApJL}}

\def\pasp{{PASP}}
\def\ssr{{Sp. Sci. Rev.}}

\def\mcg6{{MCG-6-30-15}}

\def\ltsim{\mathrel{\hbox{\rlap{\hbox{\lower4pt\hbox{$\sim$}}}\hbox{$<$}}}}
\def\gtsim{\mathrel{\hbox{\rlap{\hbox{\lower4pt\hbox{$\sim$}}}\hbox{$>$}}}}

\def\rg{{$R_{g}$}}

\begin{document}

\title[Swift X-ray/UV/optical variability of NGC5548]{Swift monitoring of NGC5548:
X-ray reprocessing and short term UV/optical variability}

\author[M$\rm^{c}$Hardy, I.M.]
{ I M M$\rm^{c}$Hardy$^{1}$, D T Cameron$^{1}$, T Dwelly$^{2,1}$, S
  Connolly$^{1}$, P Lira$^{3}$,  \and D Emmanoulopoulos$^{1}$,
J. Gelbord$^{4}$,
E. Breedt$^{5}$, P Arevalo$^{6,7}$, and P Uttley$^{8}$\\
$^{1}$ Department of Physics and Astronomy, The University, Southampton
SO17 1BJ\\
$^{2}$Max-Planck-Institut f{\"u}r extraterrestrische Physik,
Giessenbachstrasse 1, D-85748, Garching, Germany \\
$^{3}$ Departmento de Astronomia, Universidad de Chile, Camino del Observatorio 1515, Santiago, Chile\\
$^{4}$ Spectral Sciences Inc, 4 Fourth Avenue,
Burlington, MA 01803 USA\\
$^{5}$ Department of Physics, University of Warwick, Coventry CV4 7AL\\
$^{6}$ Instituto de Astrof\'isica, Facultad de F\'isica, Pontificia
Universidad Cat\'olica de Chile, 306, Santiago 22, Chile\\
$^{7}$ Instituto de F\'isica y Astronom\'ia, Facultad de Ciencias, Universidad de Valpara\'iso, Gran Breta\~na N 1111, Playa Ancha, Valpara\'iso, Chile\\
$^{8}$ Astronomical Institute `Anton Pannekoek', University of Amsterdam, Science Park 904, NL-1098 XH Amsterdam, the Netherlands
}

\date{Accepted 2014 July 22. Received 2014 July 22; in original form 2014 July 08}
\pubyear{2014}

\maketitle

\begin{abstract}
Lags measured from correlated X-ray/UV/optical monitoring of
AGN allow us to determine whether UV/optical variability is driven
by reprocessing of X-rays or X-ray variability is driven by UV/optical
seed photon variations. We present the results of the largest study to
date of the relationship between the X-ray, UV and optical variability
in an AGN with 554 observations, over a 750d
period, of the Seyfert 1 galaxy NGC5548 with Swift. There is a good
overall correlation between the X-ray and UV/optical bands,
particularly on short timescales (tens of days). These bands
lag the X-ray band with lags which are proportional to wavelength to
the power $1.23 \pm 0.31$.
This power is very close to the power (4/3) expected if
short timescale UV/optical variability is driven by reprocessing of X-rays by a
surrounding accretion disc.

The observed lags, however, are longer than expected from a standard
Shakura-Sunyaev accretion disc with X-ray heating, given the currently
accepted black hole mass and accretion rate values, but can be
explained with a slightly larger mass and accretion rate, and a
generally hotter disc.

Some long term UV/optical variations are not paralleled exactly in the
X-rays, suggesting an additional component to the UV/optical
variability arising perhaps from accretion rate perturbations propagating
inwards through the disc.

\end{abstract}

\section{Introduction}
\label{sec:intro}

\begin{figure*}
\psfig{figure=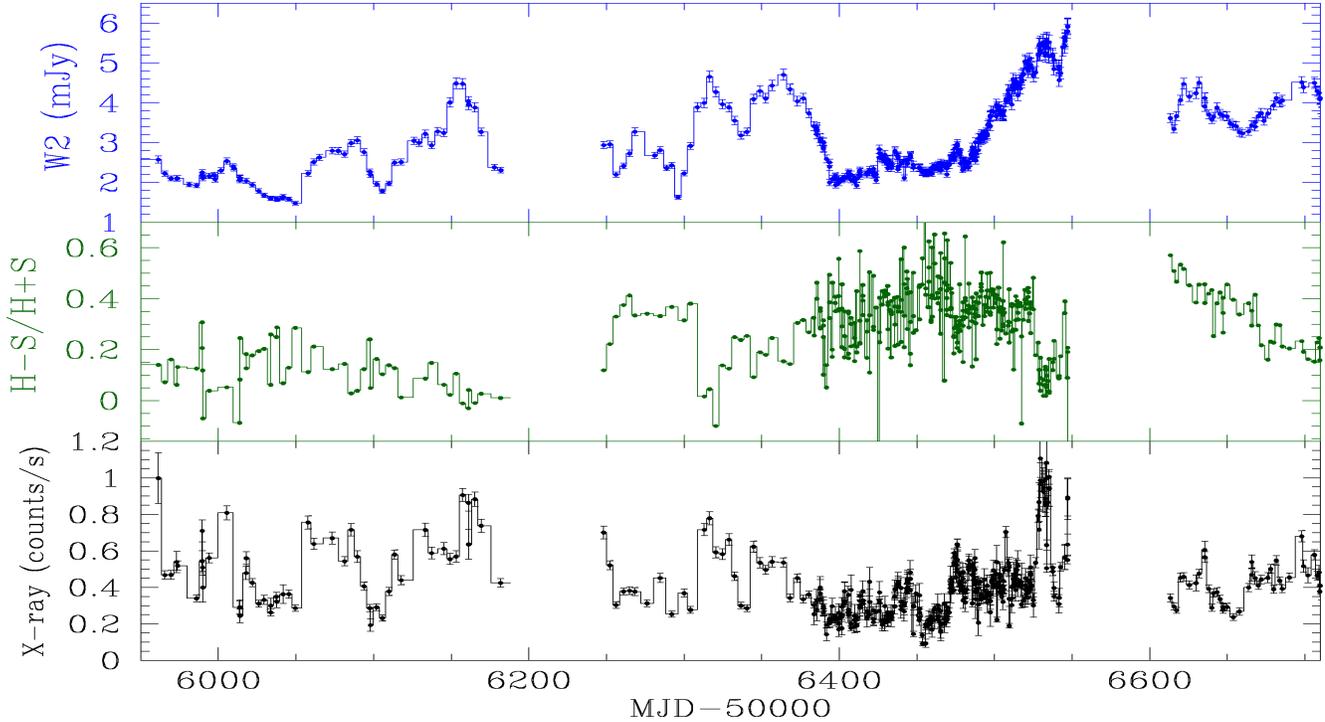,width=175mm,height=95mm,angle=0}
\vspace*{-2mm}
\caption{{\it (Bottom panel)} Long term Swift 0.5-10 keV X-ray count
  rate, {\it (Middle panel)}
  X-ray hardness ratio, where H = 2-10 keV count rate and S = 0.5-2
  keV count rate and {\it (Top Panel)}  UVW2 flux. }
\label{fig:lclong}
\end{figure*}
\vspace*{-2mm}

The origin of the UV and optical variability in AGN, and its
relationship to the X-ray variability, is still a subject of
considerable debate. A number of studies \cite[e.g.][]{uttley03_5548,
  arevalo08_2251, arevalo09, breedt09, breedt10, lira11, cameron12, cameron14,
  shappee14} have shown strong X-ray/UV or X-ray/optical correlations
on short timescales (weeks - months), with lags close to zero days,
but poorer correlations on longer timescales (months - years), usually
due to long term  UV/optical trends which are
not mirrored in the X-ray variability.
These observations suggest that different processes dominate the
UV/optical variability on different timescales. An X-ray/UV
correlation on short timescales could result if the UV/optical
emission is produced by reprocessing of X-rays by the nearby accretion
disc. The UV/optical would then lag the X-rays by the short light
travel time between the X-ray source and the disc. Alternatively, if
the X-ray variations are produced by variations in the UV seed photon
flux, produced by accretion variations in the disc at very small
radii, then the X-rays should lag the UV-optical by that same short
light travel time. Thus determining the precise lag between the X-ray
and UV-optical emission is a strong diagnostic of the origin of the
UV-optical variability.

Almost all previous studies, based mainly on a combination of X-ray
monitoring with RXTE and ground based optical monitoring, show short
($\sim1$~d) lags of the X-rays by the optical. However in no
individual case is the lag measured well enough to rule out,
unambiguously, that the optical might lead. Ground based optical
monitoring suffers from interuptions by bad weather but the Swift
observatory can provide regular
simultaneous X-ray, UV (UVW2, UVM2, UVW1) and optical (U,B,V)
monitoring, allowing measurement of wavelength dependent lags

Based on Swift observations \cite{cameron12} were able to show B-band
lagging the X-rays by less than $<45$mins in NGC4395 and
\cite{shappee14}, using Swift and ground based
observations, were able to measure interband lags in NGC2617 which
were in good agreement with a reprocessing model. Purely within the
optical bands \cite{sergeev05} and \cite{cackett07} measured lags
consistent with a reprocessing origin.

\cite{shappee14} observe for 50d with almost daily sampling
following NGC2617 in outburst. Here we report on 554 observations of
the Seyfert 1 galaxy NGC5548 over a 750d period. These
observations were largely made as a result of our own proposals but
also contain some archival data from other programs \cite[e.g.][]{kaastra14}.
Our observations were not scheduled to follow particular flares and so are
typical of the long term behaviour of NGC5548. This AGN is already known
to show a strong X-ray/V-band correlation
\cite{uttley03_5548} but the V-band lag has not yet been
precisely defined and no simultaneous UV/X-ray monitoring has yet
taken place.

In Section 2 we discuss the Swift observations and lightcurves and in
Section 3 we discuss the interband correlations and lags on both long
and short timescales. In Section 4 we compare our lag measurements to
those expected from reprocessing from a standard \cite{shakura73}
accretion disc model and in Section 5 we discuss the implications of
our results for the origin of the variability in the optical, UV and
X-ray bands in Seyfert galaxies.

\begin{figure*}
\psfig{figure=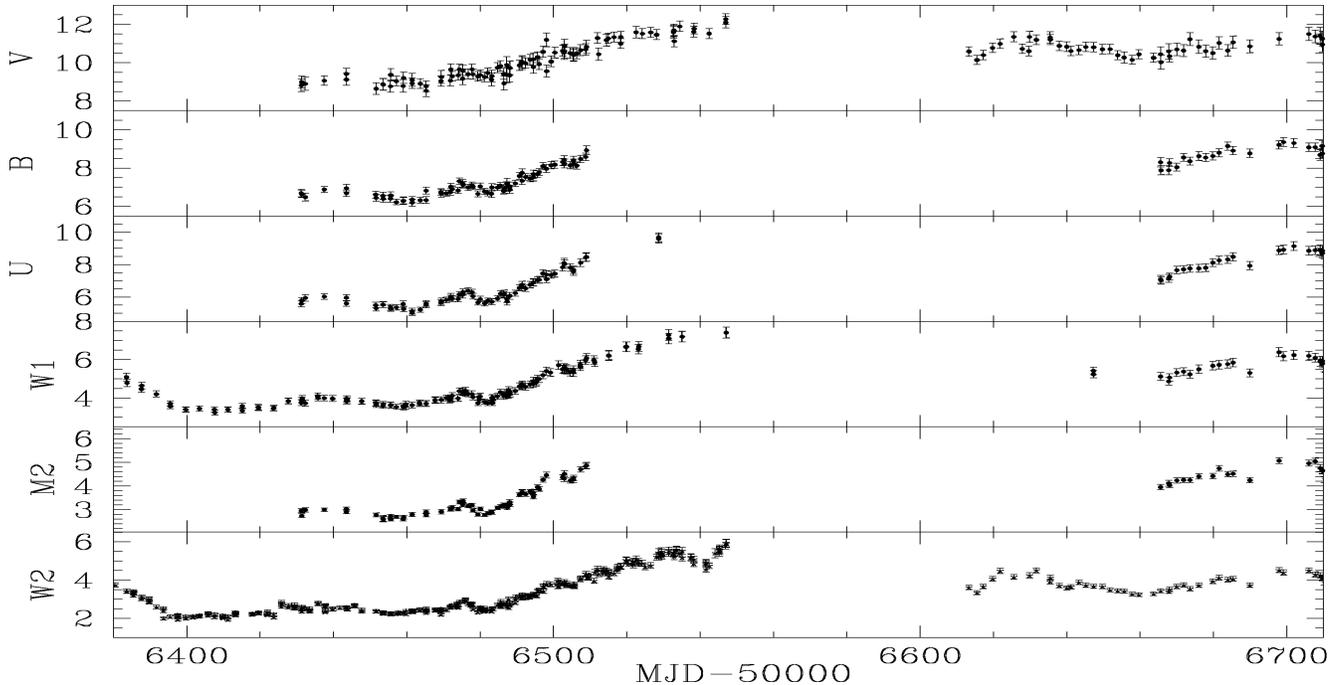,width=175mm,height=90mm,angle=0}
\vspace*{-3mm}
\caption{Multiband UVOT light curves in mJy. \protect\footnotesize
  (Colour version available in source files.) }
\label{fig:lc_6band}
\end{figure*}

\section{SWIFT Observations}

The SWIFT X-ray observations are made by the X-ray Telescope
\cite[XRT,][]{burrows05} and UV and optical observations are made by
the UV and Optical Telescope \cite[UVOT,][]{roming05}.  The XRT
observations were carried out in photon-counting (PC) mode and UVOT
observations were carried out in image mode. These data were analysed
using our own pipeline which is based upon the standard Swift analysis
tasks as described in \cite{cameron12}.  X-ray data are corrected for
the effects of vignetting and aperture losses and data obtained when
the source was located on known bad pixels, or with large flux error
($>0.15$ count s$^{-1}$) are rejected. Unless otherwise stated we use
the 0.5-10 keV X-ray band. Occasional UVOT datapoints were seen to lie
15\% or more below the local mean, usually as a result of extremely
rapid drops. All such datapoints were examined individually
and some were found to be the result of bad tracking and were
rejected. The rest occur when the image falls on
particlar areas of the detector, suggesting one or more bad
pixels. All such data were removed.

Here we consider the long, well sampled, period from MJD-50000 of 5960 to 6709
although there are a small number of earlier observations.  Dates
hereafter are in these units.  Observations, mostly of 1~ks though
sometimes of 2~ks, typically occurred every 2 days although periods of
less frequent (4d) or more frequent (1d) sampling occurred. Each
observation was usually split into 2 or more individual visits,
improving time sampling. In total 554 visits were made. After
rejection of bad data 465 X-ray measurements remain, with 300
occurring between Day 6383 and 6547. Initially we restricted our UVOT
observations to UVW2,
following guidelines to reduce filter wheel rotations.  Later we
relaxed the rotation constraint and additional filters were
used.

The UVW2 band was sampled in almost all visits and the resultant X-ray
count rate and hardness ratio (H = 2-10 keV; S = 0.5-2 keV), and UVW2
flux lightcurves, at the highest available time resolution, are shown
in Fig~\ref{fig:lclong}. \citep[][S4, show a 1~d binned version of
  this figure]{kaastra14}.
The gaps centred on 6220 and 6580 result from
Swift sun-angle constraints.  We see a generally close correspondence
between the X-ray and UVW2 flux lightcurves as noted previously
between the X-ray and V bands by \cite{uttley03_5548}.  The
correspondence on short timescales ($\sim10$d) is strong but the
amplitudes of variability on longer timescales are not always
identical, e.g.  from Day 6470 to 6550 when UVW2 shows a strong upward
trend with a much weaker trend in the X-rays.
From Day 6380 additional observations were made in additional UVOT
filters and the resultant light curves are shown in
Fig.~\ref{fig:lc_6band}. We see a close correspondence between all
UVOT bands.

\section{X-ray / UV-Optical Correlations}

\begin{figure}
\psfig{figure=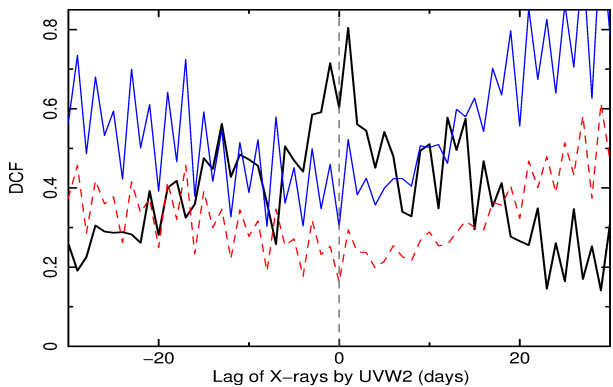,width=80mm,height=50mm,angle=0}
\vspace*{-2mm}
\caption{ Discrete cross correlation function between the X-ray
  and UVW2 lightcurves shown in
  Fig.~\protect\ref{fig:lclong}. The 95\% (dashed red) and
  99.99\% (solid thin blue) confidence levels are also shown. }
\label{fig:xw2_dcf}
\end{figure}

\begin{figure}
\psfig{figure=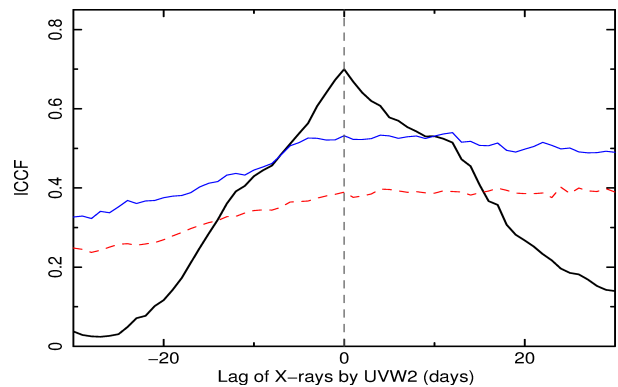,width=80mm,height=50mm,angle=0}
\caption{Interpolation cross correlation function between the X-ray
  and UVW2 lightcurves shown in Fig.~\ref{fig:lclong}. The 95\%
  (dashed red) and 99\% (dashed blue)
confidence levels are also shown.}
\label{fig:xw2_iccf}
\end{figure}

\begin{figure}
\psfig{figure=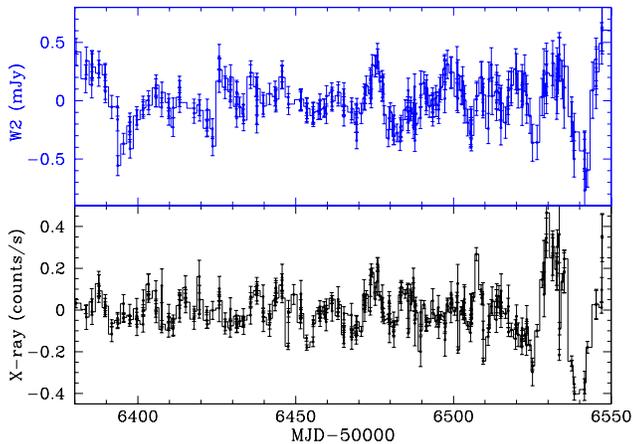,width=85mm,angle=270}
\vspace*{-4mm}
\caption{Swift 0.5-10 keV X-ray (bottom panel) and UVW2 (top panel)
  lightcurves for the intensively sampled period. The mean level from
  a 20d running boxcar has been subtracted from both light curves.
}
\label{fig:intensive}
\end{figure}

\subsection{X-ray / UVW2 correlation}

The UVW2 band is the best sampled of the UVOT bands and in
Figs~\ref{fig:xw2_dcf} and \ref{fig:xw2_iccf} we show the discrete
\cite[DCF,][]{edelson88} and interpolation \cite[ICCF,][]{gaskell86}
CCFs \citep[see also][]{white_peterson94} for the complete X-ray and
UVW2 datasets from Day 5960 to 6710.  Only the more slowly varying
UVW2 band is interpolated.  We do not interpolate over the two large
gaps but take the weighted mean ICCF of the 3 sections seen in
Fig~\ref{fig:lclong}. The N\% confidence levels are defined such that
if correlations are performed between the observed UVW2 data and
randomly simulated X-ray lightcurves with the same variability
properties as the observed data \citep[e.g.][]{summons07_phd}, only (100-N)\% of the correlations
would exceed those levels \cite[e.g. see][for more details]{breedt09}.
The confidence levels are appropriate to a single trial, ie a search
at zero lag, approximately what we are investigating here.
Both functions show a broad, but highly significant correlation,
peaking near zero lag, with the DCF favouring UVW2 lagging the
X-rays by about a day.

CCFs can be distorted when there is a long term variation in the mean
level in one light curve which is not present in the other and so it
is recommended practice to subtract a running mean \citep{welsh99}.  We
therefore subtracted a mean based on a running boxcar of width 20d
from both UVW2 and X-ray light curves. As 300 of the 465 good X-ray
datapoints lie within the intensively sampled period from 6383 to 6547
whose data dominate the measurement of short timescale lags, we
concentrate on this period.  where $\sim2$ visits per day are made. As
the separation between visits is not uniform, good sampling of
sub-daily variations is provided.  The resulting 20d mean-subtracted
UVW2 and X-ray light curves for the intensive period are shown in
Fig.~\ref{fig:intensive}.

\begin{figure}
\psfig{figure=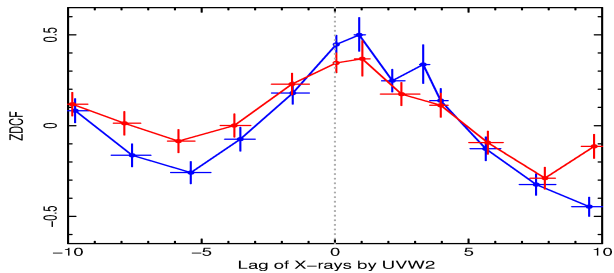,width=80mm,height=35mm,angle=0}
\caption{ZDCF between the X-ray and UVW2 lightcurves from
  Day 6383-6547. The blue
  line is the result when both X-rays and UVW2 have been mean
  subtracted using a 20d
  running boxcar (Fig.~\ref{fig:intensive}) and the red line
  is the result when only the UVW2 has been mean subtracted.
}
\label{fig:intensive_zdcf}
\end{figure}

We have calculated a variety of DCFs, ICCFs and ZDCFs
\citep[][Fig.~\ref{fig:intensive_zdcf}]{alexander13}, and all show a
UVW2 lag of $\sim$one day. Subtracting a 40d running mean
gives a similar result. 

\begin{figure}
\centering
\begin{minipage}[b]{0.45\linewidth}
\hspace*{-5mm}
\psfig{figure=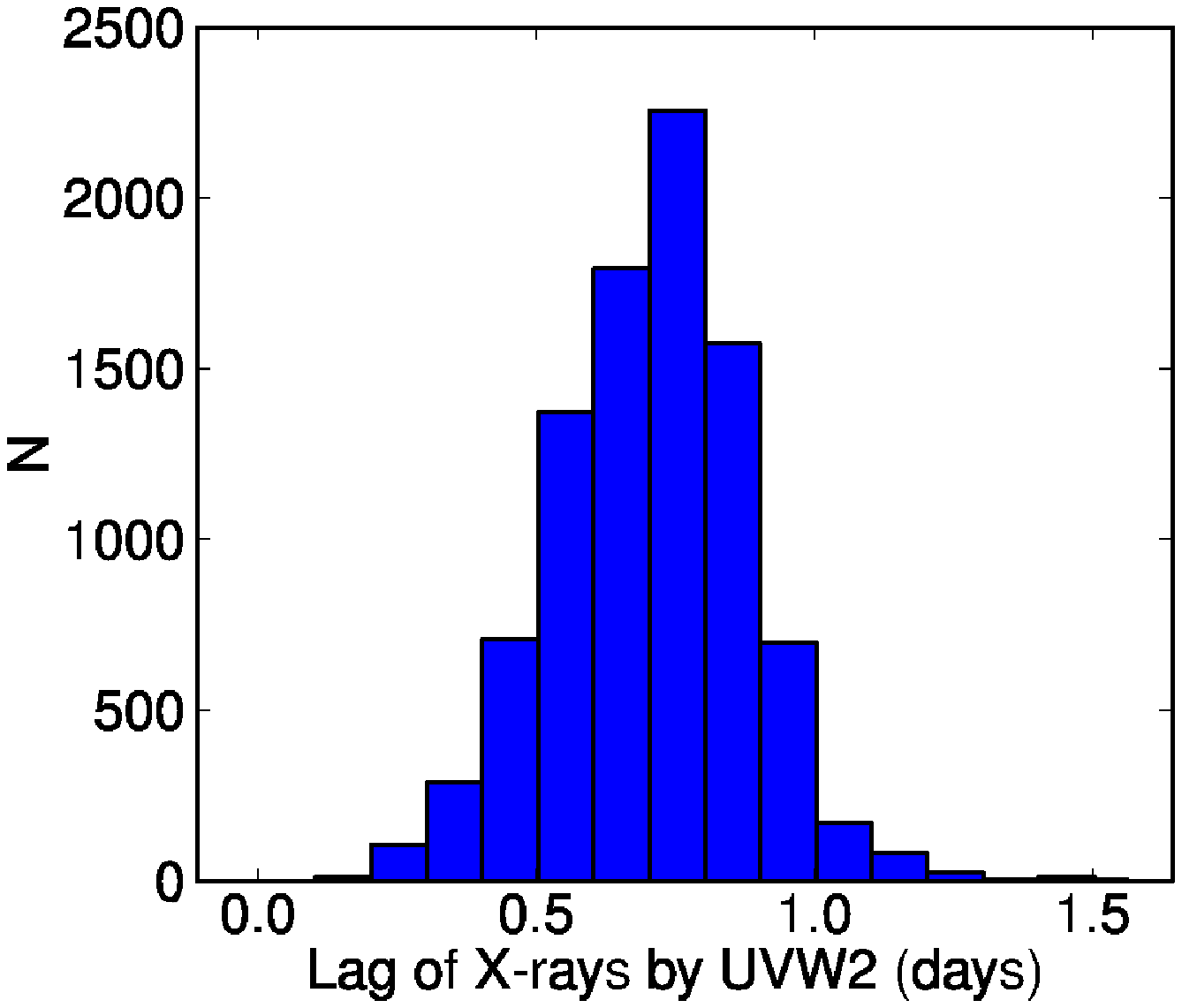,width=40mm,height=30mm,angle=0}
\end{minipage}
\begin{minipage}[b]{0.45\linewidth}
\psfig{figure=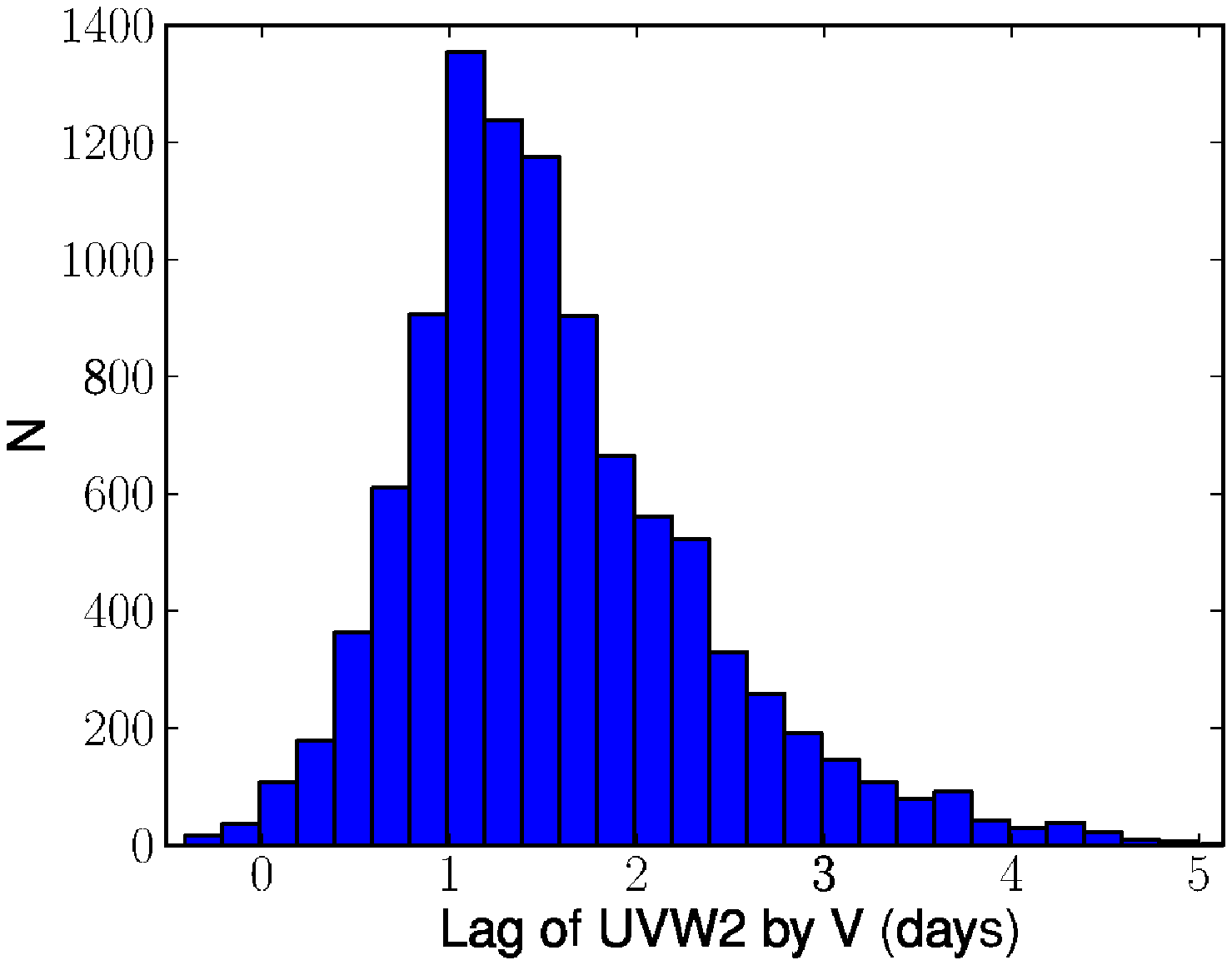,width=40mm,height=30mm,angle=0}
\end{minipage}
\caption{Lag distributions from Javelin: {\it Left Panel} UVW2 following
  X-rays using data shown in Fig.~\ref{fig:intensive}. {\it Right
    Panel} V following UVW2 using data shown in Fig.~\ref{fig:lc_6band}.
}
\label{fig:jav}
\end{figure}
To refine the lag measurement we followed \cite{shappee14}, and the
recommendation of \cite{pancoast14}, and
calculated the distribution of lags from 10000 simulations
for the period Day 6383 - 6547 
using the JAVELIN cross-correlation program \citep{zu11_javelin,zu13_javelin}. We
allowed a search range of $\pm10$d. JAVELIN assumes a similar pattern
of variability in both bands. Although the X-ray and UVW2 variability
properties will be slightly different, the lag measurement should not
be noticeably affected \citep{shappee14}.  The resulting lag
distribution is shown in Fig.~\ref{fig:jav} (left panel). The median
UVW2 lag is +0.70$^{+0.24}_ {-0.27}$d where the errors are the
standard JAVELIN 16\% and 84\% probability levels. Changing the
observation period does not noticeably change the result; eg for the
period 6248-6547 the derived lag is +0.79$^{+0.21}_ {-0.32}$d and if
we use the whole dataset from 5960 to 6710 the lag is 0.62$\pm
0.35$d.  When applied to the non-mean subtracted
  light curves, JAVELIN does not converge to the single distribution of Fig.~\ref{fig:jav} but produces a wide, multi-modal,
  distribution, presumably due to the presence of uncorrelated long
  timescale variations.
We have also calculated separately the UVW2-hard band and
UVW2-soft band lags. There is no measureable difference.

\subsection{X-ray spectral and UVW2 variations} 
\begin{figure}
\psfig{figure=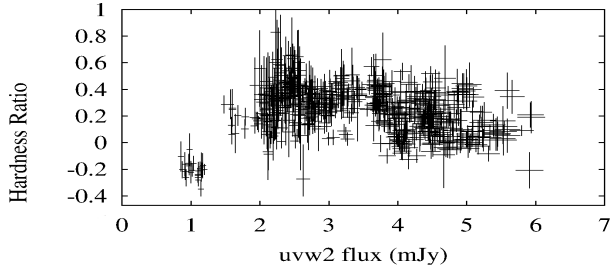,width=80mm,height=35mm,angle=0}
\vspace*{-3mm}
\caption{X-ray hardness ratio vs UVW2 flux.
}
\label{fig:hard}
\end{figure}

The X-ray spectral variations of NGC5548 have been studied
extensively, e.g. \cite{sobolewska09, kaastra14}, with the latter
showing absorption variability on different timescales, together with
intrinsic luminosity variations. In Fig.~\ref{fig:lclong} there is a
general trend for NGC5548 to become softer with increasing UVW2 and
X-ray flux \cite[as noted by][]{kaastra14}). 
In Fig.~\ref{fig:hard} we plot hardness ratio vs UVW2 flux. There is a
great deal of scatter in this relationship but the source is generally
brighter in the UVW2 band when it is softer. However the very softest
observations occur at the very lowest fluxes. Broadly similar
behaviour within the X-ray band is seen by \cite{connolly14} in
NGC1365. They explain the spectral variations with a combination of
intrinsic luminosity variations and luminosity dependent obscuration
in the context of a wind model, similar to \cite{kaastra14}.  At the
lowest fluxes only the soft unabsorbed component, scattered from the
wind, is visible.  The lack of dependence of X-ray/UVW2 lag on
  X-ray energy suggests that changing absorption does not change the
  relative paths taken by the various bands, although it may change
  the relative amplitudes, and so is not the main cause of the lags.
These spectral variations will be discussed in a future paper
(Connolly et al, in prep).
\vspace*{-5mm}

\subsection{UVOT interband lags}
All the UVOT bands show the same long term trends and so it is not
necessary to mean subtract the lightcurves. We therefore simply use
all available data, as shown in Fig.~\ref{fig:lc_6band} and calculate
the lag distributions between UVW2 and the other UVOT bands with
JAVELIN. Some distributions, such as the example shown
(Fig~\ref{fig:jav}, right panel), are slightly asymmetric. Use of the mode rather
than the median would slightly reduce the lags in some cases but the
differences are small. The X-ray, W1 and V bands have the best overlap
with W2 but we show the lags from all bands.

\subsection{Lags and X-ray reprocessing}

\begin{figure}
\psfig{figure=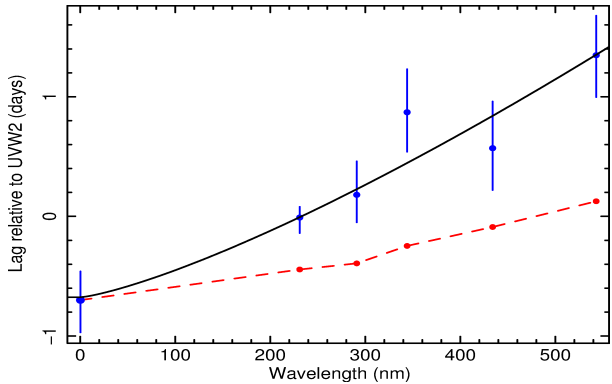,width=80mm,height=50mm,angle=0}
\vspace*{-2mm}
\caption{Lag of the X-ray and other UVOT bands relative to UVW2. Lag
  $\propto$ wavelength$^{\beta}$ where $\beta = 1.23\pm 0.31$.  The
  lower, dashed, red line is the prediction for a standard disc as described
  in the text. 
 }
\label{fig:lags}
\end{figure}

All UVOT bands lag the X-rays with lags which broadly increase with
wavelength (Fig.~\ref{fig:lags}).  We can compare the lags with the
prediction of reprocessing from a simple accretion disc where the lag
should vary as the 4/3 power of wavelength
\cite[e.g.][]{cackett07,lira11}.  Fitting a simple model of the form
$lag = A + (B\times \lambda)^{\beta}$, and assuming Gaussian
distributed errors, we find that $A =-0.70\pm 0.21$, $\beta = 1.23\pm
0.31$ and $B=3.2 \pm 0.6 \times 10^{-3}$).  The fit $\chi^{2}$
is 2.1 with 3 d.o.f. The fit goes straight
  through the X-ray point so if the lags are measured relative to the
  X-rays then, unlike \cite{shappee14}, no additional unphysical
  offset to the X-ray point is required.  The slope, $\beta$, is
however similar to that derived by \cite{shappee14} for NGC2617 ($1.18
\pm 0.33$), when including a fit offset.

This fit shows that reprocessing of X-ray emission from an accretion
disc provides a very good explanation for the short term UV/optical
variability in NGC5548.

\section{Accretion disc Modelling}

To determine whether a disc consistent with other
observed properties of NGC5548 can explain the lags we model the disc
following the prescription in \cite{lira11} \citep[c.f.][]{cameron14}
which includes X-ray as well as gravitational heating.  X-ray heating
depends on extrapolation of the observed 2-10 keV luminosity
($Lx_{2-10}$) to $\sim$0.01-500~keV, and on the disc albedo.  The
height, $H$, of the X-ray source above the disc, assuming a lamppost
geometry, and the inner disc radius, $R_{in}$, are also important
parameters.

Assuming a black hole mass $M_{BH}$ \citep[$6.7\times10^{7}$
    M$_{\odot}$,][]{bentz07} and accretion rate $\dot{m}_{E}$
  \citep[$\sim 0.03-0.04$ of Eddington][]{Vasudevan10, pounds03}, high
  X-ray heating luminosity corrected for albedo of $6 \times
  Lx_{2-10}$, implying a low albedo of 20\%, $H$=6\rg, consistent with
  X-ray source sizes measured by other methods
  \citep{chartas09,chartas12,emmanoulopoulos14} and $R_{in}$ =6\rg,
  the ISCO for a Schwarzchild black hole, we obtain the dashed line
  shown in Fig.~\ref{fig:lags}.  The predicted lags for this
  standard disc, following impulse X-ray illumination, represent when
  half of the reprocessed light has been received. The peak response
  may be even faster.

\begin{figure}
\psfig{figure=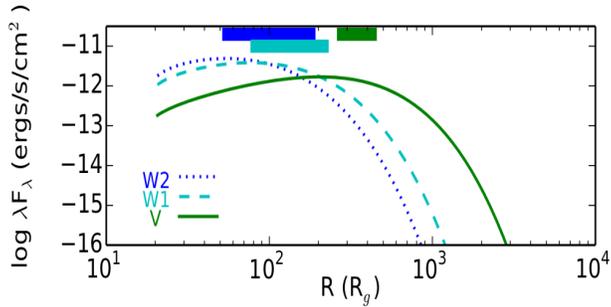,width=80mm,height=40mm,angle=0}
\caption{Disc emissivity profile in W2, W1 and V bands (lines) for the
  parameters given in the text, with measured lags, relative to the
  X-ray bands, as bars at the top.
}
\label{fig:model}
\end{figure}

To increase the predicted lags to agree with observation, in this
homogeneous disc model, we have to change the geometry (eg $H$=20\rg,
$R_{in}$ =20\rg) and require a larger $M_{BH}=10^{8}$ M$_{\odot}$) and
hotter ($\dot{m}_{E} = 0.06$) disc. Increased disc temperature is more
important here than disc flaring.  In Fig.~\ref{fig:model} we
compare the model flux distributions for the above parameters as a
function of radius in the W2, W1 and V bands, which are the best
sampled, with the radii estimated from the lags. There is reasonable
agreement.  These lags will be modelled in more detail in a future
paper (Lira et al, in prep.).

\section{Discussion}

{\it Longer than expected lags:} Although the wavelength-dependent
lags in Fig.~\ref{fig:lags} strongly support reprocessing of X-rays
are the major cause of short-timescale UV/optical variability in AGN, 
our observations imply that the
reprocessed emission comes from further from the black hole
than expected for a `standard' accretion disc. By pushing the
parameter limits we can reconcile prediction and observation but we
note that \cite{morgan10} also require a larger than expected disc to
explain their microlensing observations. They suggest a low radiative
efficiency, which is consistent with our requirement for $R_{in} \geq
20$. An alternative solution is provided by \cite{dexter11} who
propose inhomogeneous discs where the outer portions will contribute
more flux than for a uniform disc, causing the disc to appear larger.

{\it Long Timescale UVOT variability:} Short timescale X-ray/UVW2
correspondence is very good, and also generally good on long
timescales, but from Day $\sim6470$ to 6547 the rise in the UV/optical
is significantly more pronounced than in X-rays, until $\sim6525$ when
a large X-ray outburst starts.  The UV/optical rise might be
  associated with an inwardly propagating rise in accretion rate which
  eventually hits the X-ray emission region. However unless the ratio
  of disc scale height to radius is larger than the normally assumed
  value of 0.1, or the rise propagates not through the disc but
  through a corona over the disc, or the perturbation starts at small
  radius ($\sim 20$\rg), the viscous timescales from 100\rg are too
  long \citep[$\sim10$~years c.f.][]{breedt09}.

{\it The relevance of seed photon variations:}
Although it is possible, by eye, to suggest periods when the
UV-optical might lead on short timescales (eg Day 6150)
such periods are rare and may just be statistical fluctuations. On
average the X-rays lead. Although this observation is simply explained
by assuming that all X-ray variability is generated within the corona,
one might ask why variations in seed photon flux, where we would
expect the UV-optical emission to lead, appear to have little affect
on the measured lags. We suggest that the answer is provided by a
combination of relative solid angles and conservation of photons
during the X-ray scattering process. Compton scattered X-ray photons
have energies 10-100$\times$ greater than the seed photon energy. As
the disc fills a large fraction of half the sky as seen by the X-ray
source, a large fraction of the Compton scattered photons will hit the
disc, out to radii beyond the source of most of the seed photons. If
these photons are absorbed by the disc, leading to black
body emission without conservation of photon number, the resultant
number of lower energy UV-optical photons produced may exceed the
initial seed photon fluctuation. This process could also add a further
delay to the optical lightcurves, aiding agreement with observation.

\section*{Acknowledgements}
This work was supported by STFC grant ST/J001600/1. PL acknowledges
grant Fondecyt 1120328. JG gratefully acknowledges the support from
NASA under award NNH13CH61C.


\end{document}